# Are all TCFTs obtained by twisting $N{=}2$ SCFTs?[1]

José M. Figueroa-O'Farrill[2]

*Department of Physics, Queen Mary and Westfield College
Mile End Road, London E1 4NS, UK*

A large class of two-dimensional topological conformal field theories (TCFTs) are obtained by the twisting construction of Witten and Eguchi–Yang. However there seem to exist TCFTs which are not obtained in this way; for instance, TCFTs obtained from the Kazama algebra and critical string theories with generic background. We will show that by embedding the critical bosonic string into the NSR string, its TCFT can indeed be obtained by twisting a $N{=}2$ SCFT. A closer look at the construction of the $N{=}2$ superconformal algebra will show that the embedding is not essential, and this will tell us how to generalise this to other string theories. We thus conclude with the natural conjecture that *all* TCFTs have a description as topologically twisted $N{=}2$ SCFTs.

---

[1] Talk given at the Workshop on Strings, Gravity and Related Topics, held at the ICTP (Trieste, Italy) on 29-30 June, 1995.

[2] e-mail: j.m.figueroa@qmw.ac.uk

## Introduction

There seems to be a tradition in the scientific literature that whenever the title of a paper—or a talk for that matter—is in the form of a "yes-no" question, the paper or talk invariably concludes by answering the question in the negative. Alas, I will not quite break this tradition, but neither shall I follow it. Instead, I hope that by the end of this talk, I manage to convince you that the answer to the question posed in the title is probably "Yes."

To understand what this question means, we need to get a few concepts straight. I will be very brief because in the Summer School we have just heard two beautiful series of talks by Greene and Bershadsky which contained quite a lot about $N{=}2$ superconformal field theories (SCFTs) and about topological conformal field theories (TCFTs), respectively.

## What is a TCFT?

We shall not be very precise, but simply list some the gross features of any two-dimensional TCFT. We will only treat the holomorphic sector. For the purposes of this talk, a TCFT will mean a conformal field theory with the following properties:

- the energy-momentum tensor $\mathbb{T}^{\text{top}}(z)$ has zero central charge and $L_0$ acts diagonally;

- there exists a fermionic current $\mathbb{G}^+(z)$, called the BRST current, whose zero-mode $Q = \oint \mathbb{G}^+(z)$ obeys $Q^2 = 0$;

- there exists a fermionic field $\mathbb{G}^-(z)$, such that $\mathbb{T}^{\text{top}}(z) = [Q, \mathbb{G}^-(z)]$; and

- there exists a bosonic current $\mathbb{J}(z)$, whose charge $q = \oint \mathbb{J}(z)$ acts diagonally and is such that $\mathbb{G}^\pm(z)$ has charge $\pm 1$.

The space of states of the TCFT is then the BRST cohomology $H^\bullet(Q)$, which is graded by the eigenvalues of the charge $q$. Because the energy-momentum tensor $\mathbb{T}^{\text{top}}(z)$ is BRST exact, all correlation functions of BRST-invariant fields with $\mathbb{T}^{\text{top}}(z)$ insertions are identically zero. Such correlation functions are therefore topological invariants. In particular, since $\mathbb{T}^{\text{top}}(z)$ generates translations, correlation functions of BRST invariant fields are position independent. The cohomology $H^\bullet(Q)$ therefore inherits some algebraic structure from the CFT which begot it: from the normal-ordered product (or equivalently from the three-point function) it inherits a supercommutative associative multiplication, and from the two-point function it inherits an invariant metric. This makes $H^\bullet(Q)$ into a Frobenius (super)algebra. This is not all, however, and in all known examples, $H^\bullet(Q)$ has some additional structure making into a Batalin-Vilkovisky (BV) algebra. This additional structure consists of a fermionic operation $\Delta : H^\bullet \to H^{\bullet - 1}$, induced from the zero mode of $\mathbb{G}^-(z)$, obeying $\Delta^2 = 0$. By a well-known procedure [1], $\Delta$ allows us to define on $H^\bullet(Q)$ a Gerstenhaber bracket. This brief summary on the algebraic structures on a TCFT cannot possibly do justice to the excellent treatments to be found in the literature: both in the context of vertex operator algebras [2], which is closer in spirit to our approach, and in the context of operads [3].

Let us conclude this brief introductory section by recording a useful definition. For the present purposes, we will say that two TCFTs are **equivalent** if they give rise to isomorphic BV algebras.

## Topologically twisted $N=2$ SCFTs

Despite the suggestive notation, the definition of a TCFT does not imply that $\mathbb{T}^{\text{top}}(z)$, $\mathbb{G}^\pm(z)$, and $\mathbb{J}(z)$ should actually obey the OPEs of a topologically twisted $N=2$ superconformal algebra; although this is in fact the most common example.

The $N=2$ superconformal algebra is generated by fields $\mathbb{T}(z)$, $\mathbb{G}^\pm(z)$, and $\mathbb{J}(z)$ subject to the following OPEs:

$$\begin{aligned}
\mathbb{G}^\pm(z)\mathbb{G}^\pm(w) &= \text{reg} \\
\mathbb{G}^+(z)\mathbb{G}^-(w) &= \frac{d}{(z-w)^3} + \frac{\mathbb{J}(w)}{(z-w)^2} + \frac{\mathbb{T}(w) + \frac{1}{2}\partial\mathbb{J}(w)}{z-w} + \text{reg} \\
\mathbb{J}(z)\mathbb{G}^\pm(w) &= \frac{\pm \mathbb{G}^\pm(w)}{z-w} + \text{reg}
\end{aligned} \quad (1)$$

The remaining OPEs follow as a consequence of the associativity of the operator product expansion [4] and since they are standard we will not write them here.

Given the above $N=2$ superconformal algebra we can construct a TCFT in one of two ways, known as twisting [5]. We shall be focusing for definiteness on the TCFT defined by $\mathbb{T}^{\text{top}}(z) = \mathbb{T}(z) + \frac{1}{2}\partial\mathbb{J}(z)$. Then $H^\bullet(Q)$ is nothing but the chiral ring of the $N=2$ SCFT, which has a BV operator $\Delta$ induced from the zero mode of $\mathbb{G}^-(z)$. We will call a TCFT obtained in this way an $N=2$ TCFT.

The simplest example of an $N=2$ TCFT is the following: take a fermionic BC system $(b,c)$ of weights $(\lambda, 1-\lambda)$ and a bosonic BC system $(\beta,\gamma)$ of the same weights. We take as defining OPEs:

$$\beta(z)\gamma(w) = \frac{1}{z-w} \qquad b(z)c(w) = \frac{1}{z-w} \;.$$

Out of these fields we can make the following composite fields:

$$\begin{aligned}
\mathbb{G}^+_K &= b\gamma \\
\mathbb{G}^-_K &= \lambda \partial c\beta + (\lambda-1)c\partial\beta \\
\mathbb{J}_K &= (1-\lambda)bc + \lambda\beta\gamma \\
\mathbb{T}^{\text{top}}_K &= \lambda\left(\beta\partial\gamma - b\partial c\right) + (\lambda-1)\left(\partial\beta\gamma - \partial bc\right)\;,
\end{aligned} \quad (2)$$

which satisfy a topologically twisted $N=2$ SCFT. The reason why I say this is the simplest TCFT is that the chiral ring of this theory is generated by the identity, a fact that follows from the Kugo-Ojima quartet mechanism—after making a choice of "picture" of the $(\beta,\gamma)$ system. The subscript $K$ is short for "Koszul–Kugo–Ojima," and we will call such a TCFT a KKO TCFT, for short.

Let us now record a simple fact, which will be of use later: if $(\mathbb{T}^{\text{top}}, \mathbb{G}^\pm, \mathbb{J})$ and $(\mathbb{T}^{\text{top}'}, \mathbb{G}^{\pm'}, \mathbb{J}')$ define TCFTs, which need not come from twisting $N=2$ SCFTs, the combinations $(\mathbb{T}^{\text{top}} + \mathbb{T}^{\text{top}'}, \mathbb{G}^\pm + \mathbb{G}^{\pm'}, \mathbb{J} + \mathbb{J}')$ define a TCFT, which we call the tensor product of the original TCFTs. In particular, tensoring by a KKO TCFT yields a TCFT which is equivalent to the original TCFT. In other words, we may think of the KKO TCFT as the identity under the operation of taking tensor product in the space of (equivalence classes of) TCFTs.

## TCFTs not defined by twisting $N=2$ SCFTs

The above definition of a TCFT seems to allow for more general algebraic structures than the ones afforded by an $N=2$ SCFT. We will now briefly review two of them.

The search for more general algebraic structures underlying TCFTs led Kazama [6] to define a generalisation of the $N=2$ superconformal algebra, which twists to give rise to a TCFT. This algebra is generated by fields $\mathbb{T}(z)$, $\mathbb{G}^\pm(z)$, $\mathbb{J}(z)$, $\Phi(z)$, and $\mathbb{F}(z)$ subject to the following OPEs:

$$\begin{aligned}
\mathbb{G}^+(z)\mathbb{G}^+(w) &= \text{reg} \\
\mathbb{G}^+(z)\mathbb{G}^-(w) &= \frac{d}{(z-w)^3} + \frac{\mathbb{J}(w)}{(z-w)^2} + \frac{\mathbb{T}(w) + \frac{1}{2}\mathbb{J}(w)}{z-w} + \text{reg} \\
\mathbb{J}(z)\mathbb{G}^\pm(w) &= \frac{\pm\mathbb{G}^\pm(w)}{z-w} + \text{reg} \\
\mathbb{G}^-(z)\mathbb{G}^-(w) &= \frac{-2\mathbb{F}(w)}{z-w} + \text{reg} \\
\mathbb{G}^+(z)\Phi(w) &= \frac{\mathbb{F}(w)}{z-w} + \text{reg} \\
\mathbb{J}(z)\Phi(w) &= \frac{-3\Phi(w)}{z-w} + \text{reg}\;.
\end{aligned} \quad (3)$$

As in the $N=2$ superconformal algebra, there are more nonzero OPEs but they are uniquely characterised by these [7]. We call this algebra a Kazama algebra. If we twist $\mathbb{T}^{\text{top}}(z) = \mathbb{T}(z) + \frac{1}{2}\partial\mathbb{J}(z)$, then we have a TCFT.

The Kazama algebra appears naturally in the context of the $G/G$ gauged WZW model [8] and, more generally, there exists a construction in terms of Manin pairs [7]. Notice that if we put $\Phi(z) = \mathbb{F}(z) = 0$, then the Kazama algebra reduces to an $N=2$ superconformal algebra; but for nonvanishing $\Phi(z)$ and $\mathbb{F}(z)$, the Kazama algebra seems to give rise to novel TCFTs.

Another class of TCFTs which are not defined by twisting $N=2$ SCFTs are string theories. We will consider, as an example, the bosonic string with background an arbitray CFT defined by an energy-momentum tensor $T_M(z)$ with central charge $c_M{=}26$, and with fermionic ghosts $(b,c)$ of weights $(2,-1)$. The relevant composite fields defining the TCFT of the bosonic string are

$$\begin{aligned}\mathbb{G}^+_{N=0} &= T_M c + bc\partial c \\ \mathbb{G}^-_{N=0} &= b \\ \mathbb{J}_{N=0} &= -bc \\ \mathbb{T}^{\text{top}}_{N=0} &= T_M - 2b\partial c - \partial bc \;,\end{aligned} \qquad (4)$$

Notice that the OPE of the BRST current $\mathbb{G}^+_{N=0}(z)$ with itself is not regular. Since the BRST current is defined only up to a total derivative, one may try to "improve" it to cancel the singular part of the OPE; but it is easily shown that with the fields that we have available $(T_M, c, b)$ this is impossible. Hence the resulting TCFT does not seem to be a $N{=}2$ TCFT. Of course, for special $T_M(z)$—for example in the case of the noncritical strings, when $T_M(z)$ has a Liouville part—there may be ways to improve the BRST current and the $U(1)$ current $\mathbb{J}(z)$ to make a $N{=}2$ superconformal algebra, as was found in [**9**]; but this is not the case for a general background. Similar conclusions hold for other string theories: NSR, $\mathsf{W}_3$,...

These two examples tempt us to conclude, at least naively, that there are TCFTs which do not come from twisting $N{=}2$ SCFTs.

String embeddings

The first hint that the above conclusion may be misleading comes from the well-known string embedding of the bosonic string in the NSR string [**10**]; which shows that the same TCFT (in this case the bosonic string) may be described by more than one conformal field theory. Let us review this briefly.

Let $(T_M, c_M{=}26)$ be a bosonic string background, and let $(\tilde{b}, \tilde{c})$ be a fermionic BC system of weights $(\frac{3}{2}, -\frac{1}{2})$. Out of these fields we can construct the following composite fields:

$$\begin{aligned}T_{N=1} &= T_M - \tfrac{3}{2}\tilde{b}\partial\tilde{c} - \tfrac{1}{2}\partial\tilde{b}\tilde{c} + \tfrac{1}{2}\partial^2(\tilde{c}\partial\tilde{c}) \\ G_{N=1} &= \tilde{b} + T_M \tilde{c} + \tilde{b}\tilde{c}\partial\tilde{c} + \tfrac{5}{2}\partial^2\tilde{c}\end{aligned} \qquad (5)$$

which obey an $N{=}1$ superconformal algebra with central charge $c{=}15$. This makes $(T_{N=1}, G_{N=1})$ into an NSR string background, albeit of a very special kind. To analyse the TCFT arising from this NSR string we introduce the superconformal ghost systems: a fermionic $(b,c)$ system of conformal weights $(2,-1)$ and a bosonic $(\beta,\gamma)$ system of weights $(\frac{3}{2},-\frac{1}{2})$. We can now write down the generators $(\mathbb{T}^{\text{top}}, \mathbb{G}^\pm, \mathbb{J})$ which characterise the TCFT resulting from this NSR string. Normally we would not expect to be able to improve these fields to make them obey an $N{=}2$ superconformal algebra, but because of the form of this NSR string background, this is indeed possible. In fact, let us define

$$\begin{aligned}\mathbb{G}^+_{N=1} &= cT_{N=1} + \gamma G_{N=1} - b\partial cc - b\gamma^2 + b\beta\partial\gamma - \tfrac{1}{2}\partial c\beta\gamma + \partial X \\ \mathbb{G}^-_{N=1} &= b \\ \mathbb{J}_{N=1} &= -bc + \tfrac{1}{2}\beta\gamma + \tfrac{1}{2}\tilde{b}\tilde{c} + \tilde{c}b\gamma + \tfrac{1}{2}\partial\tilde{c}\tilde{c}\beta\gamma - \tfrac{1}{4}\partial^2\tilde{c}\tilde{c} \\ \mathbb{T}^{\text{top}}_{N=1} &= T_{N=1} - 2b\partial c - \partial bc + \tfrac{3}{2}\beta\partial\gamma + \tfrac{1}{2}\partial\beta\gamma \;,\end{aligned} \qquad (6)$$

where

$$X = \tfrac{1}{2}\tilde{b}c\tilde{c} + \tfrac{1}{2}\partial c - \tfrac{1}{2}\beta\tilde{c}c\partial\tilde{c}\gamma + \tfrac{1}{2}\tilde{c}\beta\gamma^2 - bc\gamma\tilde{c} - \partial\tilde{c}\gamma - \tfrac{1}{4}c\tilde{c}\partial^2\tilde{c} \;.$$

It is not hard to prove that the above fields define a topologically twisted $N{=}2$ TCFT. Since the TCFT resulting from this NSR string is equivalent to the one resulting from the bosonic string defined by $T_M(z)$, we conclude that *any* bosonic string is equivalent (as a TCFT) to one resulting from twisting an $N{=}2$ SCFT.

A more natural interpretation

As a matter of fact, the embedding of the bosonic string in the NSR string is not really necessary to understand what is actually happening here. Indeed, let us consider the following current:

$$\mathbb{R} = -\tilde{c}b\gamma - \tilde{c}c\partial\beta + \tfrac{3}{2}\tilde{c}\partial c\beta + \tfrac{1}{2}\tilde{c}\partial\tilde{c}bc - \tfrac{1}{4}\tilde{c}\partial\tilde{c}\beta\gamma \qquad (7)$$

with zero mode $R = \oint \mathbb{R}(z)$. It turns out that $R$ is ad-nilpotent, so that we may conjugate by it. Doing so we find:

$$\begin{aligned}e^R \mathbb{G}^+_{N=1} e^{-R} &= cT_M + bc\partial c + \tilde{b}\gamma + \partial Y \\ e^R \mathbb{G}^-_{N=1} e^{-R} &= b + \tfrac{1}{2}\tilde{c}\partial\beta + \tfrac{3}{2}\partial\tilde{c}\beta \\ e^R \mathbb{J}_{N=1} e^{-R} &= -bc + \tfrac{1}{2}\beta\gamma + \tfrac{1}{2}\tilde{b}\tilde{c} + \partial(c\tilde{c}\beta) \\ e^R \mathbb{T}^{\text{top}}_{N=1} e^{-R} &= T_M - 2b\partial c - \partial bc + \tfrac{3}{2}\left(\beta\partial\gamma - \tilde{b}\partial\tilde{c}\right) + \tfrac{1}{2}\left(\partial\beta\gamma - \partial\tilde{b}\tilde{c}\right) \;,\end{aligned} \qquad (8)$$

where

$$Y = \tilde{b}c\tilde{c} + c\beta\gamma - \beta\tilde{c}c\partial c + \tfrac{3}{2}\partial c \;.$$

Since conjugation is an automorphism of the OPEs, the above fields still satisfy a twisted $N{=}2$ superconformal algebra.

Notice that the above expressions for the $N{=}2$ generators are reminiscent of the tensor of product of two TCFTs: the bosonic string with background $T_M(z)$ and a KKO TCFT with fields $(\tilde{b}, \tilde{c}, \beta, \gamma)$ with $\lambda = \frac{3}{2}$. Indeed, we see that $e^R \mathbb{G}^-_{N=1} e^{-R}$ and $e^R \mathbb{T}^{\text{top}}_{N=1} e^{-R}$ are precisely the tensor product expressions, whereas $e^R \mathbb{G}^+_{N=1} e^{-R}$ is an "improvement" of the tensor product expression by a total derivative, which does not alter the zero mode. Finally, $e^R \mathbb{J}_{N=1} e^{-R}$ has been deformed the most: being "improved" not just by a total derivative, but also by a term $\tilde{b}\tilde{c} - \beta\gamma$. The fact that the KKO TCFT by which we tensor has weight $\frac{3}{2}$ is the only remnant of the embedding of the bosonic string into the NSR string. But in fact, even this is inessential, and we can tensor the bosonic string by a KKO TCFT of arbitrary weight and recover a twisted $N{=}2$ superconformal algebra. Indeed, the relevant $N{=}2$ generators are now given by

$$\begin{aligned}
\mathbb{G}^+ &= \mathbb{G}^+_{N=0} + \mathbb{G}^+_K + \partial Y \\
\mathbb{G}^- &= \mathbb{G}^-_{N=0} + \mathbb{G}^-_K \\
\mathbb{J} &= \mathbb{J}_{N=0} + \mathbb{J}_K + (\tilde{b}\tilde{c} - \beta\gamma) + \partial(c\tilde{c}\beta) \\
\mathbb{T}^{\text{top}} &= \mathbb{T}^{\text{top}}_{N=0} + \mathbb{T}^{\text{top}}_K ,
\end{aligned} \qquad (9)$$

where $Y(z)$ is the same field defined above and where the generators of the KKO TCFT can be read off from (2) *mutatis mutandis* and those for the bosonic string from (4).

Conclusion

This interpretation of the $N{=}2$ generators now makes no reference to the embedding of the bosonic string into the NSR string, and therefore has more hopes to generalise. In fact, it does; and one can show that tensoring the NSR string and the $W_3$ string by a KKO TCFT allows us to construct an $N{=}2$ superconformal algebra whose chiral ring is isomorphic as a BV algebra to the corresponding string theory. (These results will appear elsewhere.) For the Kazama algebra, this fact follows from formulas obtained in [**7**] in the process of coupling the Kazama algebra to topological gravity—the relevant KKO theory being the semi-infinite Weil complex of the Virasoro algebra, in this case. I should mention also that the fact that the NSR string is equivalent to an $N{=}2$ TCFT was already shown in [**11**], exploiting the embedding (see the first reference of [**10**]) of the NSR string into the $N{=}2$ string.

These results essentially take care of all the TCFTs I know, whence in the absence of counterexamples, I dare conclude this talk with the following conjecture:

*Every TCFT is equivalent to an N=2 TCFT*

ACKNOWLEDGEMENTS

I am grateful to the overseers for organising a most enjoyable Summer School and for the opportunity to present these results at the Workshop. It is a pleasure to thank Sonia Stanciu for her hospitality and for enjoyable discussions. I am also grateful to Neil Marcus for making me aware of [**11**].

The work described in this talk was carried out at the University of North Carolina in Chapel Hill while I was visiting (for longer than originally expected) the Departments of Mathematics and Physics. I would like to thank Louise Dolan and Jim Stasheff for the invitation and for their warm hospitality. Very special thanks go to Takashi Kimura ("a friend in need...") for many enlightening conversations on this and other topics, for his patience, for taping "The Simpsons", and for putting up with me for so long without complaining. Finally, I would like to express my sincere thanks to Louise Dolan for her timely intervention; without it, I might still be in Chapel Hill.